\documentclass[sigconf]{acmart} 

\AtBeginDocument{%
  \providecommand\BibTeX{{%
    \normalfont B\kern-0.5em{\scshape i\kern-0.25em b}\kern-0.8em\TeX}}}

\copyrightyear{2022}
\acmYear{2022}
\setcopyright{acmcopyright}\acmConference[ICSE-SEIS'22]{Software Engineering in Society}{May 21--29, 2022}{Pittsburgh, PA, USA}
\acmBooktitle{Software Engineering in Society (ICSE-SEIS'22), May 21--29, 2022, Pittsburgh, PA, USA}
\acmPrice{15.00}
\acmDOI{10.1145/3510458.3513021}
\acmISBN{978-1-4503-9227-3/22/05}

\usepackage{adjustbox}
\usepackage{xspace}
\usepackage{tabularx}
\usepackage{colortbl}
\usepackage{subfig}
\usepackage{multirow}
\usepackage{booktabs}
\usepackage{listings}
\usepackage{scratch3}
\usepackage{tikz}
\usepackage[capitalise]{cleveref}

\usepackage{dingbat}
\usepackage{marvosym}
\usepackage{balance}

\setscratch{scale=.60}

\newcommand\definetool[2]{\newcommand{#1}{{\textsc{#2}}\xspace}}
\definetool{\Scratch}{Scratch}
\definetool{\leila}{LeILa}
\definetool{\whisker}{Whisker}
\definetool{\litterbox}{LitterBox}
\definetool{\bastet}{Bastet}
\definetool{\scratchblocks}{scratchblocks}

\colorlet{punct}{red!60!black}
\definecolor{background}{HTML}{EEEEEE}
\definecolor{delim}{RGB}{20,105,176}
\colorlet{numb}{magenta!60!black}

\lstdefinelanguage{json}{
    basicstyle=\normalfont\ttfamily,
    numbers=left,
    numberstyle=\scriptsize,
    stepnumber=1,
    numbersep=8pt,
    showstringspaces=false,
    breaklines=true,
    frame=lines,
    backgroundcolor=\color{background},
    literate=
     *{0}{{{\color{numb}0}}}{1}
      {1}{{{\color{numb}1}}}{1}
      {2}{{{\color{numb}2}}}{1}
      {3}{{{\color{numb}3}}}{1}
      {4}{{{\color{numb}4}}}{1}
      {5}{{{\color{numb}5}}}{1}
      {6}{{{\color{numb}6}}}{1}
      {7}{{{\color{numb}7}}}{1}
      {8}{{{\color{numb}8}}}{1}
      {9}{{{\color{numb}9}}}{1}
      {:}{{{\color{punct}{:}}}}{1}
      {,}{{{\color{punct}{,}}}}{1}
      {\{}{{{\color{delim}{\{}}}}{1}
      {\}}{{{\color{delim}{\}}}}}{1}
      {[}{{{\color{delim}{[}}}}{1}
      {]}{{{\color{delim}{]}}}}{1},
}

\usepackage{fbox}
\newcommand{\summary}[2]{
        \vspace{3mm}
        \noindent
        \fbox{%
            \parbox{.97\linewidth}{%
                    \textbf{#1 Summary.}
                #2
            }%
        }%
}%

\usepackage[colorinlistoftodos,prependcaption,textsize=tiny]{todonotes}

\newcommand\spiral{%
  \begin{tikzpicture}[y=.009ex,x=.009ex,yscale=-1]
    \path[fill] (74.2500,379.5469) -- (62.4375,373.3594) .. controls
      (83.4375,365.2031) and (99.6563,356.1563) .. (111.0938,346.2188) .. controls
      (122.5313,336.2813) and (131.0625,324.9375) .. (136.6875,312.1875) .. controls
      (142.3125,299.4375) and (145.1250,284.4375) .. (145.1250,267.1875) .. controls
      (145.1250,256.6875) and (141.9375,247.5000) .. (135.5625,239.6250) .. controls
      (129.1875,231.7500) and (121.7813,226.3125) .. (113.3438,223.3125) .. controls
      (104.9063,220.3125) and (95.6250,218.8125) .. (85.5000,218.8125) .. controls
      (75.3750,218.8125) and (66.0937,220.3125) .. (57.6562,223.3125) .. controls
      (49.2187,226.3125) and (41.8125,231.7500) .. (35.4375,239.6250) .. controls
      (29.0625,247.5000) and (25.8750,256.6875) .. (25.8750,267.1875) .. controls
      (25.8750,277.6875) and (27.3750,286.9687) .. (30.3750,295.0312) .. controls
      (33.3750,303.0937) and (38.1563,308.9063) .. (44.7188,312.4688) .. controls
      (51.2813,316.0313) and (59.2500,317.8125) .. (68.6250,317.8125) .. controls
      (73.6875,317.8125) and (77.5782,316.7813) .. (80.2969,314.7188) .. controls
      (83.0156,312.6563) and (85.2187,309.7500) .. (86.9062,306.0000) .. controls
      (88.5937,302.2500) and (89.4375,297.3750) .. (89.4375,291.3750) .. controls
      (89.0625,288.3750) and (88.4063,285.6563) .. (87.4688,283.2188) .. controls
      (86.5313,280.7813) and (85.0313,278.9063) .. (82.9688,277.5938) .. controls
      (80.9063,276.2813) and (78.3750,275.6250) .. (75.3750,275.6250) .. controls
      (72.9375,275.6250) and (70.5469,277.1250) .. (68.2031,280.1250) .. controls
      (65.8594,283.1250) and (63.7500,288.4219) .. (61.8750,296.0156) --
      (57.6562,295.8750) .. controls (55.5937,295.8750) and (54.0937,295.4063) ..
      (53.1562,294.4688) .. controls (52.2187,293.5313) and (51.7500,292.5000) ..
      (51.7500,291.3750) .. controls (51.7500,286.5000) and (52.6875,281.7187) ..
      (54.5625,277.0312) .. controls (56.4375,272.3437) and (59.2500,268.5000) ..
      (63.0000,265.5000) .. controls (66.7500,262.5000) and (70.8750,261.0000) ..
      (75.3750,261.0000) .. controls (80.6250,261.0000) and (85.4063,262.1250) ..
      (89.7188,264.3750) .. controls (94.0313,266.6250) and (97.4063,270.0937) ..
      (99.8438,274.7812) .. controls (102.2813,279.4687) and (103.5000,285.0000) ..
      (103.5000,291.3750) .. controls (103.5000,299.6250) and (102.3750,306.5625) ..
      (100.1250,312.1875) .. controls (97.8750,317.8125) and (94.0313,322.5937) ..
      (88.5938,326.5312) .. controls (83.1563,330.4687) and (76.5000,332.4375) ..
      (68.6250,332.4375) .. controls (57.3750,332.4375) and (47.0625,329.8125) ..
      (37.6875,324.5625) .. controls (28.3125,319.3125) and (21.5625,311.5313) ..
      (17.4375,301.2188) .. controls (13.3125,290.9063) and (11.2500,279.5625) ..
      (11.2500,267.1875) .. controls (11.2500,254.0625) and (14.2500,241.9687) ..
      (20.2500,230.9062) .. controls (26.2500,219.8437) and (35.1563,211.3125) ..
      (46.9688,205.3125) .. controls (58.7813,199.3125) and (71.6250,196.3125) ..
      (85.5000,196.3125) .. controls (99.3750,196.3125) and (112.2187,199.3125) ..
      (124.0312,205.3125) .. controls (135.8437,211.3125) and (144.7500,219.8437) ..
      (150.7500,230.9062) .. controls (156.7500,241.9687) and (159.7500,254.0625) ..
      (159.7500,267.1875) .. controls (159.7500,287.4375) and (156.5625,304.7813) ..
      (150.1875,319.2188) .. controls (143.8125,333.6563) and (134.0625,346.1250) ..
      (120.9375,356.6250) .. controls (107.8125,367.1250) and (92.2500,374.7656) ..
      (74.2500,379.5469) -- cycle;
  \end{tikzpicture}%
}

\begin{document}

\title{\Scratch as Social Network: \\ Topic Modeling and Sentiment Analysis in \Scratch Projects}

\author{Isabella Graßl}
\email{isabella.grassl@uni-passau.de}
\affiliation{%
  \institution{University of Passau}
  \state{Passau}
  \country{Germany}
}

\author{Gordon Fraser}
\email{gordon.fraser@uni-passau.de}
\affiliation{%
  \institution{University of Passau}
  \city{Passau}
  \country{Germany}
}

\begin{abstract}
  Societal matters like the \textit{Black Lives Matter} (BLM) movement
  influence software engineering, as the recent debate on replacing
  certain discriminatory terms such as \textit{whitelist/blacklist}
  has shown. Identifying relevant and trending societal matters is
  important, and often done for traditional social media channels such as Twitter. In this
  paper we explore whether this type of analysis can also be used for
  introspection of the software world, by looking at the thriving
  scene of \Scratch programmers.
  The educational programming language \Scratch is not only used for
  teaching programming concepts, but also offers a platform for young
  programmers to express and share their creativity on any topics of
  relevance.
  By automatically analyzing titles and project comments in a dataset of
106.032 \Scratch projects, we explore which topics are common in the \Scratch community, whether socially relevant events are reflected, and how the sentiment in the comments discussing these topics is.
  It turns out that the
  diversity of topics within the \Scratch projects makes the analysis
  process challenging. Our results nevertheless show that topics from pop and net culture are present, and even recent societal events
  such as the Covid-19 pandemic or BLM are to some extent reflected in \Scratch. The tone in the comments is mostly positive with catchy youth language.
  Hence, despite the challenges, \Scratch projects can be studied in the same way as social networks, which opens up new possibilities to improve our understanding of the behavior and motivation of novice programmers.
\end{abstract}



\maketitle

{\bf Lay Abstract:} Analyzing the content and emotions within a social network  provides insights on how software engineers communicate and what they talk about. The \Scratch programming environment is extremely popular with young, learning programmers. In this paper we investigate whether social network analysis in terms of automatically identifying topics and sentiments can also be applied to \Scratch. We analyze the topics of the projects that young programmers create and share, and determine the tone of their conversations. Although the process turns out to be technically challenging, we encounter pop and net culture references but also societal matters, and the overall tone of communication is positive.

\section{Introduction}
\label{sec:intro}
Societal issues such as the \textit{Black Lives Matter} movement (BLM)
have an impact also on software engineering (SE), for example with the
proposal to replace discriminatory terms such as
\textit{whitelist/blacklist} and \textit{master/slave}. Analysis of 
such societal events is therefore relevant
for SE as these might serve as a catalyst to question information technology conditions and for adapting contemporary patterns to a new discourse system~\cite{baruawhat2014, huinfluence2016}.
Research
investigating societal events typically focuses on social media
analyses that use text mining to extract semantic information from posts~\cite{carney2016all, manikonda2018twitter, abd2020top, ghasiya2021}. Can
we also learn about societal events from SE?

A thriving subfield in SE is the educational
programming environment \Scratch~\cite{maloneyscratch2010}, and its
millions of predominantely young users\footnote{https://scratch.mit.edu/statistics/}. 
 Prior research has
investigated \Scratch programs and 
its
users~\cite{aivaloglouhow2016, fieldsyouth2017}. However, the \Scratch ecosystem offers more than just code:
There is also surrounding textual information in terms of project
titles and users interacting by commenting and liking each
other's projects~\cite{velasqueznovice2014, morenoleonexaminin2016, grassl2021data}. This raises the question whether the analysis of social
networks~\cite{curiskis2020evaluation, wangpublic2020, ghasiya2021} can be transferred to the
analysis of \Scratch projects. A central challenge here is that \Scratch not only provides predominantly very short textual information, but above all that this is written mainly by children and teenagers. 

The aim of this paper is to identify to what extent text mining is possible in \Scratch projects and their comments, and whether socially relevant events are reflected. 
The first research question is:

\smallskip
\noindent
\textbf{RQ1: } \textit{What topics can be extracted from \Scratch projects?}
\smallskip

\noindent We conduct a first time application of automated text analysis of \Scratch projects, using the title of a project, in analogy to news headlines and Twitter hashtags~\cite{ghasiya2021}. 
We use Top2Vec~\cite{angelov2020}, a state of the art method from the field of machine learning and natural language processing for this purpose. 

%
%
The interfacial structure of the projects resembles that of a social network---with the possibility of commenting, sharing and
liking. Increased offensive behavior like hatespeech is reported for comments in traditional social networks~\cite{chetty2018hate,watanabe2018hate, weimann2020research}. 
To determine whether such behavior also exists in \Scratch comments and how semantic fields 
relate to the sentiment, our second question is: 
 
\smallskip
\noindent
\textbf{RQ2: } \textit{What is the tone of the \Scratch comments?}
\smallskip

\noindent We use sentiment analysis, which also has its origins in analysis of social networks  and recommendation systems. We analyze the project comments with the multi-class sentiment tool VADER~\cite{hutto2014vader}.


Our results show that \Scratch projects contain net and pop cultural references, but societal topics such as the Covid-19 pandemic are also referenced.
The majority of comments of the projects are positive and the users express their interest in the projects, but again we find some indicators that political events like the BLM movement are present.
With these insights, this paper lays the foundation for further interdisciplinary studies in \Scratch as a social network.
It also identifies new challenges and provides new opportunities for the application of machine learning in programming education, where text analysis is challenging due to the short, domain-independent language used by young people.
\section{Background and Related Work}
\label{sec:background}


In this paper, we bring together concepts from analyses of social networks and the
 community of young programmers in \Scratch.

\subsection{Social Network Analysis in SE}
Although analyses concerning socio-cultural issues represent a relatively young branch of research in SE, there are already some research reports on the content and structured analysis of \textit{social} networks like GitHub or Stackoverflow, similar to analysis in traditional social media~\cite{kristoufek2013, livas2018}. 
In particular, the identification of trending topics, which activities lead to better and faster reputation scores, as well as the effect of gender has been the focus of prior research~\cite{bosubuilding2013, baruawhat2014, huinfluence2016, maygender2019}.
Furthermore, language identification on Stackoverflow revealed that a consistency between the tags provided by users and the classification with a speech recognition tool is often not given~\cite{dietrichman2019}. 
%
%
Similarly, the impact and benefits of sentiment analysis for SE especially in OSS like GitHub repositories have been recently discussed \cite{guzmansentiment2014, linSentiment2018, novielliBenchmark2018, novielliSentiment2019}. Even a cross-linking study between GitHub and Twitter has recently been conducted~\cite{fang2020need}.
Identifying affective and social factors is relevant as they impact product and collaboration quality, productivity, and employee satisfaction.


\subsection{The \Scratch Programming Environment}

\Scratch is one of the most popular introductory programming environments. Shared projects can be commented on, and
include symbols for \textit{Love-it} ($\heartsuit$), \textit{Favourite}~($\star$), \textit{Remix} (\spiral), which is similar to  sharing in traditional social networks.
Looking ``inside'' a \Scratch project reveals the code of the underlying figures (\emph{Sprites}) and the environment (\emph{Stage}) of the program. Scripts are created by visually arranging blocks representing programming instructions.

The popularity of \Scratch has raised attention in research, especially in the fields of computer science education and SE.
There have been qualitative studies, e.g., regarding the digital competence of
young people and their handling of digital media by means of stories in
\Scratch~\cite{resnickmothers2012}, to encourage them to develop their own
projects and not just consume them. Some prior research also exists on the
application of machine learning in \Scratch, such as using a predictive
analysis to determine whether comments in a project are
project-related~\cite{velasqueznovice2014}, using latent class analysis to
investigate the use of programming concepts~\cite{fieldsyouth2017} or using
topic modeling to identify gender-dependent topics within
projects~\cite{grassl2021data}.
This is fundamental to better understand how children program in \Scratch in order to improve the learning environment and thus ensure long-term interest and sustainable learning outcomes---especially for underrepresented groups in SE.


\section{Method}
\label{sec:methodology}

\subsection{Dataset}
We randomly sampled publicly shared \Scratch projects using the REST API provided by the \Scratch website. 
We excluded remixes, as well as projects that contain less than ten blocks, ten views and ten loves, as such small and unpopular projects contain little information, but add noise.
We retrieved 124.160 projects, created in the period from 26.04.2007 to 18.08.2021. After preprocessing and restricting to projects created on or after 01.01.2019 to focus on recent trends, the final dataset consists of 106.032 projects for the topic analysis.
For the sentiment analysis, we included only projects which contained more than ten comments and therefore, from the original 124.160 projects only 21.786 projects remained, which was again reduced after the same preprocessing steps. The final dataset for the sentiment analysis consists of 16.816 projects.
Since we apply textual analysis, we excluded images and audio files. However, these could be included in future research, which we support by providing our analysis source code as open source.\footnote{https://github.com/se2p/semantic-analysis.git}

\subsection{Data Analysis}

\subsubsection{Preprocessing}
As with any text analysis, several preprocessing steps are required on the data. We parsed the text into \textit{tokens} (here single words), normalized to lower case and removed punctuation as well as stop words defined in the NLTK library, as they generally have no deeper or sporadic meaning, but distort the results. We also removed any characters which are not in the given ASCII range from 0 to 122, but made no filtering of English words, because the accuracy of the filters is very poor and generates many false-positives. However, the pre-trained model of Top2Vec is multilingual.
Since the inflectional form of a word does not provide relevant additional semantic information, lemmatisation with the popular Wordnet-Lemmatiser\footnote{https://www.nltk.org} is used to reduce terms to their lemma, where the frequency of the original terms is preserved.
Furthermore, we removed customized stop words relating to the project type such as ``game'', ``animation'', ``platformer'' as well as domain-specific terms such as ``sprite'' or ``remix''.
After the preprocessing, each sample (\Scratch project) contains several features---the terms of the title for the topic analysis and the terms of the comments for the sentiment analysis---which serve as input for the models.

\subsubsection{RQ1: Topic Analysis}
The title as basic feature represents the input parameter while the topic
represents the target parameter of the model. One advantage of Top2Vec is that
it automatically generates the number of topics itself and, unlike alternative
approaches, does not require choosing the number beforehand. Each topic from
the model is associated with a number of generated keywords that add more
contextual understanding of the projects' semantics.

\subsubsection{RQ2: Sentiment Analysis} 
To identify the tone 
of the projects, we use multi-class
sentiment analysis. To find the most accurate tool, two independent researchers manually classified two comments from each of 500 randomly selected projects from the dataset, hence a set of 1000 randomly selected comments. The comments were labelled into three basic moods (positive, neutral,  negative) and compared to the classification of SentiStrength\footnote{http://sentistrength.wlv.ac.uk/}, VADER\footnote{https://github.com/cjhutto/vaderSentiment} and Stanford CoreNLP\footnote{https://stanfordnlp.github.io/CoreNLP/}. SentiStrength's F1 score (0.743) is comparable to VADER (0.740), with Stanford CoreNLP (0.687) scoring lower. Since VADER is open source and has the advantage of a compound score, it was selected over SentiStrength for further analysis.

 VADER 
 is specifically adapted to sentiments expressed in social media and provides the percentage by which a text is rated positive,
negative, or neutral. In addition, a compound score is
calculated by adding the valence scores, 
 adjusting
them according to the rules, and then normalizing them 
 between -1 (most
extreme negative) and +1 (most extreme positive). Based on prior work~\cite{hutto2014vader} the threshold for positive sentiment is > 0.05
and for negative sentiment < -0.05. To assess the overall sentiment of a project,
we consider the average per comment.
The most common terms over all projects are visualized by word clouds to determine semantic fields.

\subsection{Threats to Validity}
The analysis with Top2Vec looks at the words with the highest co-occurrence. If there are important words in the proximity, the word becomes more important, but this does not have to be the case in children's projects and one should therefore also look at individual important terms with keyword analysis, tf-idf or more advanced neural networks~\cite{bhat2020}. Besides the classification into three basic moods with VADER, there are also several shades in between that enhance the understanding, e.g., anger, fun or sadness. SentiStrength-SE or Senti4SD, which are adapted specifically for SE might provide another perspective on the content; however, we explicitly decided against such one as our input is from children without reference to SE terminology.
Concerning external validity, the results of the topic analysis may not generalize to other scenarios or networks in SE such as GitHub. 
We considered all projects with a minimum of ten blocks as well as for the sentiment analysis with a minimum of ten project comments because they introduced too much noise into the data after one experiment. Nevertheless, even from these small projects interesting insights could be gained, e.g., showing self-painted backgrounds. Further filtering steps might improve the data, e.g., by excluding buggy projects.

\section{Results}
\subsection{RQ1: Topic Analysis}
\label{sec:rq1}

\begin{table*}[tb]
\centering
\footnotesize
\caption{\label{tab:topics}Automatically generated topics from the Top2Vec model with their keywords and project statistics.}
\begin{tabular}{llp{13cm}r}
\toprule
{ID} &     label  &      keywords (score) &  \# pro. \\
\midrule
0     &           inconsistency            &                  un (0.57) idk (0.56) way (0.54) ya (0.53) le (0.52) one (0.52) dude (0.51) au (0.5) dont (0.5) en (0.5) &           29552 \\
1     &  proper names &covid (0.52) sharkyshar (0.51) pfp (0.51) stickmin (0.5) tvokids (0.5) funkin (0.44) oyunu (0.42) numberblock (0.41) gacha (0.39)  &              1062 \\
5     &     animals&             cat (0.81) kitty (0.74) chat (0.63) pet (0.44) animal (0.41) mouse (0.4) bunny (0.39) dog (0.38) hedgehog (0.38) monkey (0.38) &              632 \\
22    &    pop culture (games) &  mario (0.87) pokemon (0.45) kirby (0.43) minecraft (0.41) pong (0.39) sonic (0.38) spongebob (0.38) lego (0.37) tycoon (0.37)  &              346 \\
28    &    music&           music (0.85) musical (0.62) song (0.59) rap (0.52) playlist (0.49) band (0.49) sings (0.47) rock (0.46) piano (0.44) dance (0.44) &               262 \\
43    &    pop culture (lit./film)&       potter (0.78) harry (0.64) pokemon (0.41) disney (0.37) avatar (0.34) magic (0.34) maze (0.33) cake (0.31) book (0.3) minecraft (0.3) &       212 \\
51    &    fantasy (universe) &         space (0.83) rocket (0.51) moon (0.48) alien (0.46) galaxy (0.46) planet (0.41) earth (0.4) fighter (0.39) fight (0.39) wing (0.39) &               198 \\
87    &  fantasy (gloomy) & dragon (0.85) pokemon (0.45) godzilla (0.44) wing (0.43) dinosaur (0.42) dungeon (0.42) monkey (0.42) snake (0.41) rocket (0.41)  &                147 \\
90    &  holidays &  christmas (0.84) santa (0.57) merry (0.54) winter (0.48) halloween (0.46) birthday (0.42) gift (0.4) party (0.36) potter (0.35) candy (0.34) &               145 \\
99    &    fantasy (shiny) &        rainbow (0.84) balloon (0.41) sky (0.4) unicorn (0.4) neon (0.4) purple (0.39) cloud (0.37) dragon (0.36) shark (0.35) pink (0.35) &              136 \\
\bottomrule
\end{tabular}
\end{table*}

To determine the topics of the \Scratch projects, a topic analysis was performed using Top2Vec. The model generated 1,441 topics with associated keywords. Due to the large number of generated topics and because an automatic evaluation of such unsupervised machine learning approaches is not possible, we reviewed only the 100 most frequent topics, i.e., the topics to which the most projects belong, manually. 
We manually classified these 100 topics into 10 types of topics, which are illustrated in \cref{tab:topics} with examples, and will be elaborated in the following\footnote{The complete list of generated topics as well as the dataset and the analysis are made available for replications at https://github.com/se2p/semantic-analysis.}. The ID references the rank of the topic in terms of how common this topic is.
There were also many topics that contained a combination of different topics, but did not form a coherent topic themselves, and therefore were not discussed further, but might be attractive for further studies.

The first topic in \cref{tab:topics}, labelled \textit{inconsistency}, corresponds to the most projects by far,
captured by keywords that are unrelated filler words like
\emph{one}, \emph{bit} or \emph{ok}. There are also abbreviations like \emph{idk} [i don't know] or
youth language like \emph{dude}. The model seems to assign all projects that are not associative to this topic. This might be due to the common use case for \Scratch as the students are just messing around to try out some programming
concepts in a class and therefore do not devote much attention to the titles.

The second type of topic also contains some incoherent terms
(interestingly also the term \textit{covid}), which are all to some extent \textit{proper names}. However, this topic seems to
consist of independent concepts that have no common context:
\textit{Sharkyshar} is a popular YouTube channel as well as a popular \Scratch
user. \textit{PFP} is the abbreviation for profile pic or an icon that you have
as your profile picture, while \textit{stickmini} is a video game and
\textit{tvokids} is a Canadian children's programming television network.
Therefore, this topic is an example of the model combining independent terms
without context. 
Nevertheless, this indicates that the
extraordinary event of the Covid-19 pandemic is being processed in some form in
the users' projects. So, similar to social
networks~\cite{curiskis2020evaluation, wangpublic2020, ghasiya2021}, children
in a programming environment are also responding to societal events. To what
extent this really influences programming and how it is represented should be
the subject of future work.

Some topics are related to the children's lifeworld.
One semantic field consists of different animals (Topic 5), which are
often used as characters in \Scratch, since the mascot of \Scratch is also a
cat. In \Scratch there are many different animals as characters and often
children use their pets as models for their avatars in the program.
Topic 28, \textit{music}, deals with songs, lyrics, playlist and musicals as well as different music genres such as rap and rock. Here, too, the \Scratch environment lends itself to this topic, since there are various stages, dancing figures and a dedicated block category sound, where a lot of sounds are available and you can also upload your own sounds. In addition, \textit{music} is a specific project type on the \Scratch website.
In addition, common everyday life scenarios such as holidays and celebrations are popular among the topics, which are represented by the terms \textit{Christmas}, \textit{Halloween} or \textit{birthday parties} (Topic 90). Such events seem to appeal to children, as other studies confirmed the use of \Scratch projects as ``gifts'' for Mother's Day~\cite{richardblind2016}. In particular, \Scratch reinforces all possible ways of being creative with their starter projects\footnote{https://scratch.mit.edu/starter-projects} on the website, where celebrations and parties are an appealing topic for newbies.

Many topics deal with references to net and pop culture, reflected by terms such as \textit{Pokemon}, \textit{Kirby}, \textit{Minecraft}, \textit{Sonic},  which all refer to popular computer games (Topic 22). This implies that children transfer their favorite free time activities in a programming setting. Besides, pop culture references from literature and film, especially \textit{Harry Potter} and \textit{Disney}, are evident (Topic 43). Gaining insights of such preferred references from other media could be beneficial to design programming tasks focused on this.


Abstract and fantasy worlds are part of popular topics, such as the universe (Topic 51), rather gloomy fantasy worlds with fantasy creatures such as dragons or godzilla and dinosaurs (Topic 87). Similarly, rather shiny fantasy worlds (Topic 99), characterized by rainbows, balloons and unicorns, exist. These  worlds show that the users' interests are diverse and it would therefore be intriguing to determine whether patterns can be derived between preferred topics and other factors such as gender, age or origin. In particular, the goal of further work should be to generalize this topic analysis and to link the topics with the project statistics and users in order to gain further insights into the interests of the \Scratch users.

\summary{RQ1}{Topics from net and pop culture can be found in \Scratch projects. In addition, the exceptional societal event, the Covid-19 pandemic, was also reflected in the projects.}
\subsection{RQ2: Sentiment Analysis}
\label{sec:rq3}

\begin{figure*}[tb]
	\centering
	\subfloat[\label{fig:negative} Terms of negative comments.]{\includegraphics[width=0.32\linewidth]{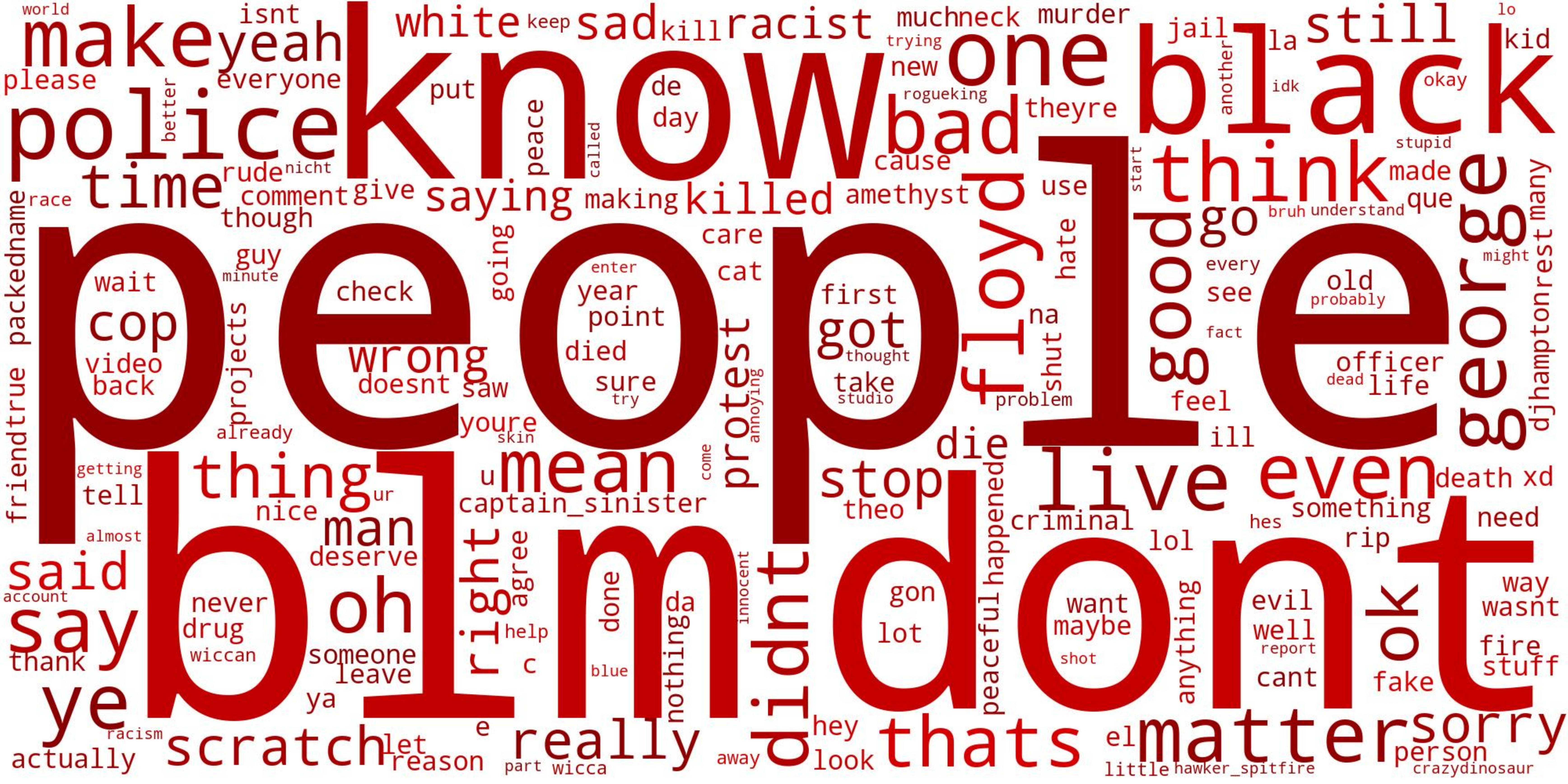}}
	\hfill
	\subfloat[\label{fig:neutral} Terms of neutral comments.]{\includegraphics[width=0.32\linewidth]{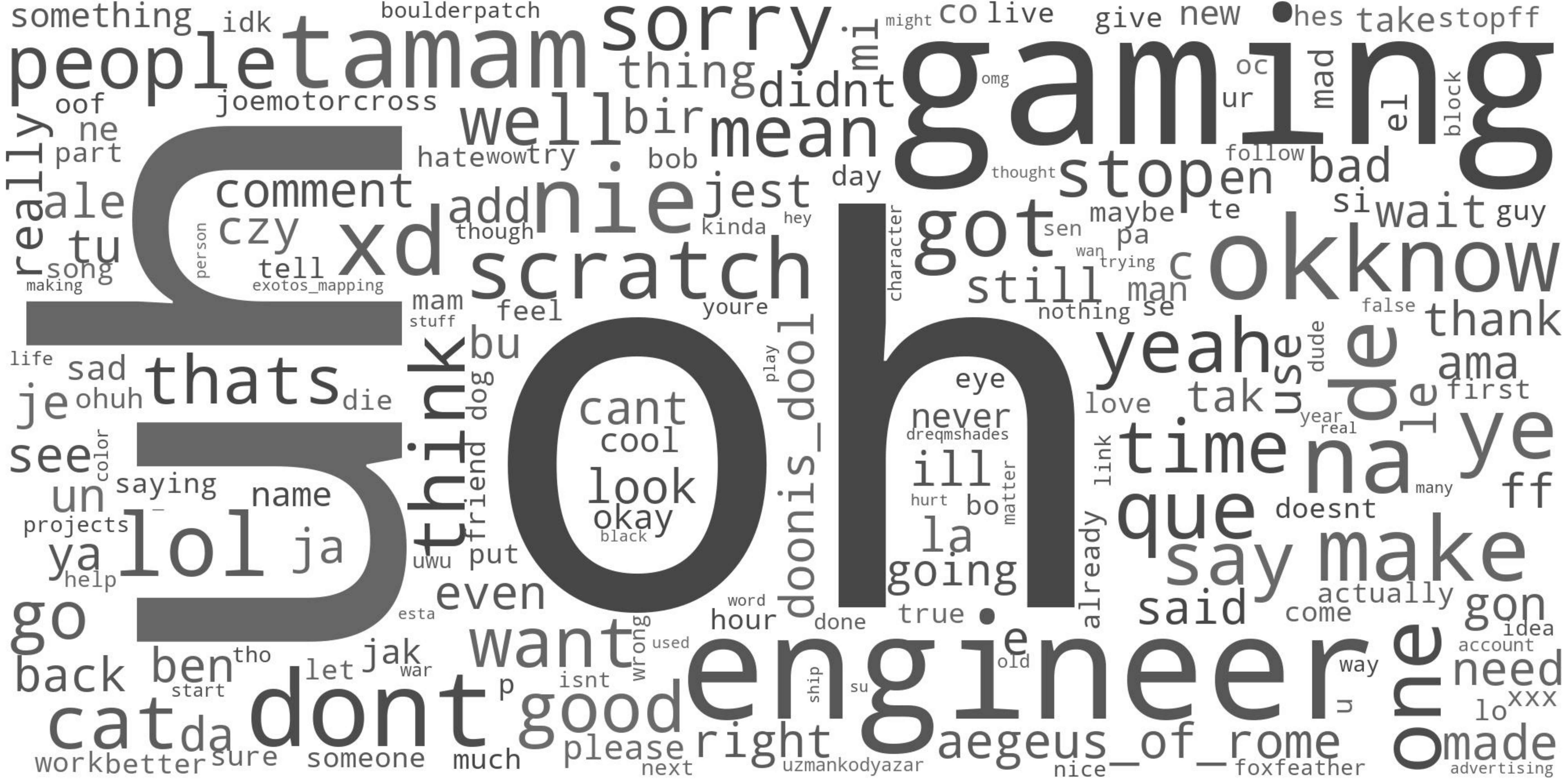}}
	\hfill
	\subfloat[\label{fig:positive} Terms of positive comments.]{\includegraphics[width=0.32\linewidth]{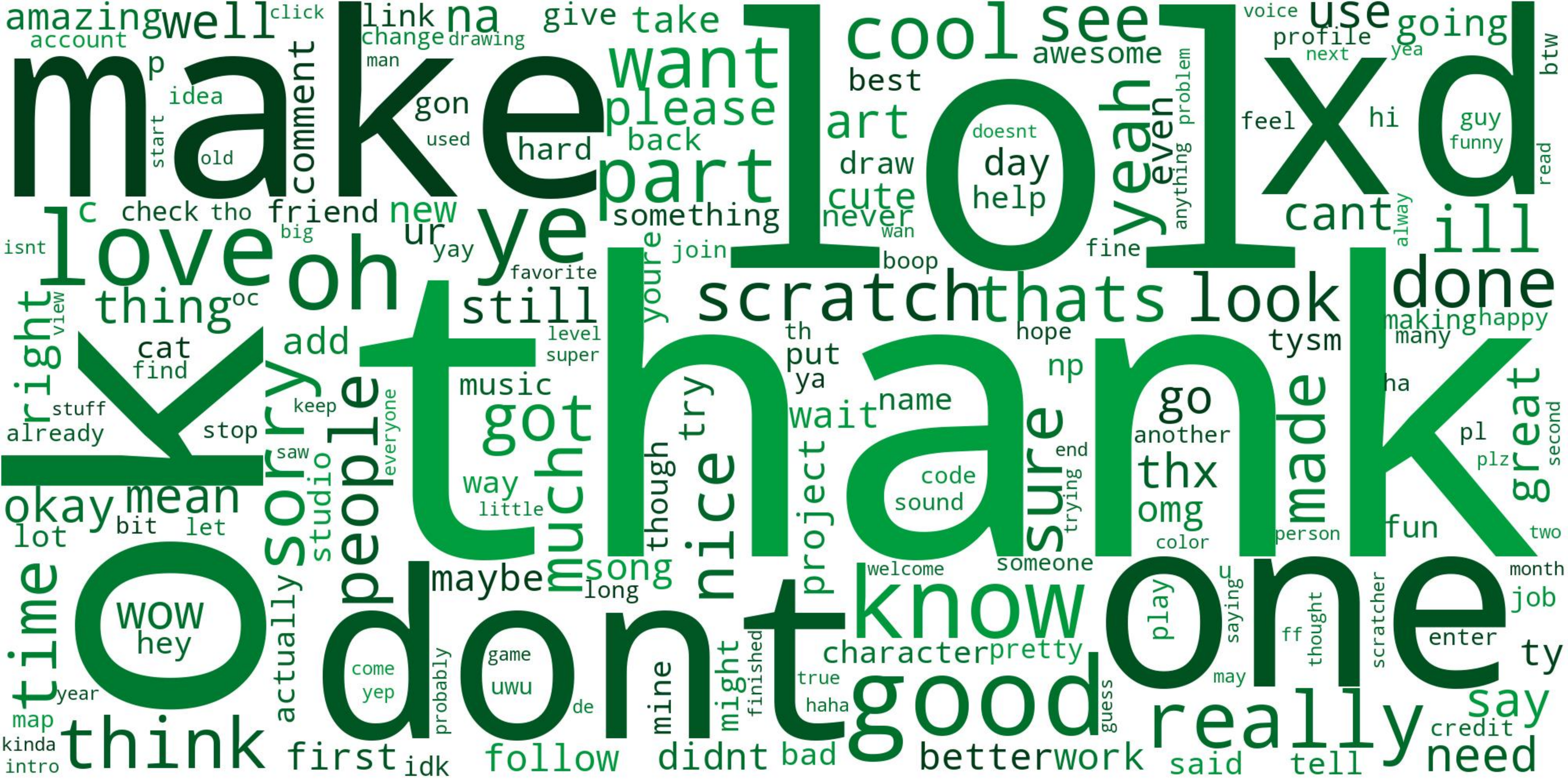}}
	\caption{Most common terms in \Scratch comments classified to their sentiment. The size of the term indicates its frequency.}
	\label{fig:sentimentwordclouds}
\end{figure*}

Overall, the comments from the projects consist of 49.362 nouns, 3.751 verbs, 3.064 adjectives and 1.005 adverbs. \Cref{fig:sentimentwordclouds} shows the most common terms in \Scratch project comments according to their sentiment. 
In total, there are 14.434 positive projects, 1.933 neutral projects and 449 negative projects.
It is clearly evident that comments of a positive nature clearly outweigh negative comments. 
In contrast to other social networks such as Instagram~\cite{hosseinmardi2015detection}, Twitter~\cite{watanabe2018hate} or YouTube~\cite{mathew2019thou} in which cyber-bullying, hatespeech as well as radicalizations have already established themselves ~\cite{chetty2018hate}, \Scratch still seems to be a friendly platform. Similar to the recent social platform TikTok~\cite{weimann2020research}, this positive communication may be due to the fact that (1) the age group of users is rather low and (2) there are less controversial or political debates than, e.g., on Twitter. 

The positive comments are dominated by the term \textit{thank}. 
The expression of thanks is a sign of
appreciation---either from users to the project creator or from project
creators in response to comments on their projects. In addition, there are
strong positive words such as \emph{love}, \emph{best}, \emph{great}, \emph{awesome} or \emph{pretty}, \emph{cute} or
\emph{nice}. These terms in combination with the other terms like \emph{character}, \emph{work}, \emph{job}
or \emph{done} result in phrases that praise good performance or a pleasant
implementation of the program. In particular, acronyms like \textit{lol}
[laughing out loud] or \textit{xd} [smiley face] dominate also the positive
project comments. Here, the specific speech situation is unclear, whether in
its euphoric or ironic meaning. In either case these acronyms exist for
identification as well as differentiation from other groups. A prerequisite for
this is to understand the content of the term and to put it into
context---\textit{lol}, for example, is a sign for colloquial language of
younger persons, i.e., youth slang~\cite{noauthorlangenscheidt2011}.

The word field of negatively connotated comments is
dominated by the word \textit{people}, which cannot be inferred in the
context. However, among the negative comments, it is striking that the acronym
\textit{BLM} [black lives matter] as well as the terms \textit{George},
\textit{Flyod}, \textit{police}, \textit{racist} and \textit{black} can be found, referring to the event which
was the reason for the movement in the USA. This is a strong indicator that
societal and political relevant events were reflected in the projects
or at least mentioned in the comments. As a consequence, \Scratch might not be \textit{just} a
programming platform, but might have the potential of being a social media
channel, which should be a aim for further research~\cite{carney2016all}.
In addition, there are also strong
negative words like \textit{kill(ed)} or \textit{die}, which
could be read in the context of the BLM movement, but might be also related to the project type: Especially in
games, it is often mentioned in the comments that the game players died very
quickly, such as \textit{I already died at level 2}. The use of the terms
\textit{death} or \textit{kill} usually has no pejorative connotation in this
context.

The neutral comments contain expressions of astonishment like \textit{uh} and \textit{oh}, which have to be interpreted depending on the context, which this analysis does not provide. In addition, the terms \emph{gaming} and \emph{engineer} stand out, which on the one hand refer to the popular project type game and on the other hand to the type of producer of a project. Overall, neutral words appear to be of little significance.

We observe in \cref{fig:sentimentwordclouds} that words such as \textit{make}, \textit{one}, \textit{dont}, or \textit{know} overlap between sentiments, implying that they are not uniquely assignable or simply context-independent. Since these words are attributed so little specific information content, they might have to be ignored for further studies. Often terms also denote ambiguity, but this seems to be less evident in the projects. 
Overall, it seems that positive comments tend to focus on the project in general and its implementation (\textit{good job}), while negative comments tend to focus on the content level of the project. This, however,  would need to be verified in further studies. 

\summary{RQ2}{The tone in the comments is mostly positive and mostly specific terminology from youth language is used. Also, political events like the BLM movement are referenced.}
\section{Discussion}

We identified topics from pop culture as well as from society in \Scratch projects. The tone of the discussions around these projects was mostly positive. The intersection of those two aspects are not only relevant for research purposes, but also for educational use.


\subsection{Topics and Sentiment}
Since we discovered a rather unexpected variety of topics (RQ1), and also controversial topics like the BLM movement in the comments (RQ2), we hypothesize that socially and political controversial topics might be discussed more controversially in the comments in \Scratch. 
In this context, likes, loves and remixes might be helpful to determine which project topics are particularly popular and the extent to which the community approves or disapproves these topics.
Our findings in detecting cultural and political topics provide the foundation for investigating in detail whether \Scratch also represents a political network where different political positions are present and debated.


In this respect, since we know that gender-specific preferred themes exist~\cite{grassl2021data} and since we have identified such topics with \textit{shiny} and \textit{gloomy} fantasy worlds (RQ1) in this paper as well, a social network analysis in combination with our approach would help to determine whether certain user groups concentrate on certain topics. In the current discourse on socialization in SE, identifying \textit{influencers} of the \Scratch community, their range of topics, their commenting behavior would provide insights into the relationship between social and programming behavior~\cite{morenoleonexaminin2016}.


\subsection{Educational Impact}
Similar to social media, it is necessary to understand in \Scratch with what topics and attitudes children and adolescents are engaged or confronted. 
Since we identified a broad range of diverse topics in RQ1, we can learn what motivations there actually are for children to program in \Scratch~\cite{aivaloglouhow2016, grassl2021data} in order to ensure that children, especially underrepresented groups such as girls, are initially attracted to introductory courses in the first place and also to follow up on children's continued long-term interest after they have been introduced to \Scratch~\cite{fieldsyouth2017}. 

As the results of RQ2 imply, there is the potential to incorporate \Scratch into the classroom besides CS also for cross-curricular education, e.g., when discussing politically relevant topics in the projects and also highlighting different perspectives in the commentary narratives.
With regard to digital literacy, which is becoming increasingly important, the children can be made aware of how social media operate using the \Scratch platform and be familiarized with real social networks. On the one hand, \Scratch can be used to playfully demonstrate the strengths of social media, e.g., in the exchange of controversial topics, but also where it is important to be careful, e.g., with negative comments.

Since we observed some phrases like \textit{good job} (RQ2), it may be possible  to determine from the comments whether a project receives a lot of positive or negative feedback and loves and likes primarily because of its topic, or because of its implementation; this would be of relevance from a computer science education point of view. In this context, it is also relevant whether the comments are purely affective, which may be substantive in terms of mood, but do not provide any information about the nature and manner of the project, i.e., whether there is also constructive criticism and suggestions for improvement of the implementation in the comments.

\section{Conclusions and Future Work}
\label{conclusions}

In this paper we explored whether social network analysis can be applied to
\Scratch. We particularly found topics from net and pop culture, but there are
also references to societal topics, such as the Covid-19 pandemic. This
supports the interpretation of \Scratch not just as a programming language, but
also as a segment of society, which provides many opportunities for further
interdisciplinary studies. For example, likes, loves and remixes could be
analyzed with the aim of determining which topics are particularly liked or
shared.
In an educational context, the correlations between project themes and the
degree of experience of the user might be relevant for introductory courses in
computer science didactics. User group-specific preferences might be identified
in order to enable children, especially girls, to be better supported on an
educational level with regard to their enthusiasm towards programming.
Since the topic modeling was challenging, a supervised machine learning
approach might support more concrete topic classification and predicting
thematically trends on the platform in future research.

\begin{acks}
This work is supported by the Federal Ministry of Education and Research
through project ``primary::programming'' (01JA2021) as
part of the ``Qualitätsoffensive Lehrerbildung'', a joint initiative of the
Federal Government and the Länder. The authors are responsible for the content
of this publication.
\end{acks}

\balance

\bibliographystyle{ACM-Reference-Format}
\bibliography{references}

\end{document}